\documentclass[prd,preprintnumbers,12pt]{revtex4}
\usepackage{epsfig} 
\usepackage{graphicx}

\newcommand{\be}{\begin{equation}}
\newcommand{\dd}{\displaystyle}
\newcommand{\ee}{\end{equation}}
\newcommand{\bea}{\begin{eqnarray}}
\newcommand{\eea}{\end{eqnarray}}

\newcommand{\nn}{\nonumber}
\newcommand{\de}{\partial}

\begin{document}
\title{Playing with fermion couplings in Higgsless models}

\author{R. Casalbuoni, S. De Curtis, D. Dolce and D. Dominici}
\affiliation{Department of Physics, University of Florence, and
INFN, Florence, Italy}

\date{\today}

\begin{abstract}
\noindent
We discuss the fermion couplings in a four dimensional $SU(2)$ linear
moose  model
by allowing for direct  couplings between the left-handed fermions on the boundary
and the gauge fields in the internal sites. This is realized
 by means of a product of non linear $\sigma$-model scalar fields which, in the continuum  limit,
is equivalent to a
Wilson line. The effect of
these new non local couplings is a contribution to
 the $\epsilon_3$ parameter which can be  of  opposite sign with respect to the one coming from the gauge fields along the string. Therefore, with some fine tuning, it
is possible to satisfy the constraints from the electroweak data.
\end{abstract}
\pacs{xxxxxxx} \maketitle
\newcommand{\f}[2]{\frac{#1}{#2}}
\def\to{\rightarrow}
\def\ptl{\partial}
\def\beq{\begin{equation}}
\def\eeq{\end{equation}}
\def\bea{\begin{eqnarray}}
\def\eea{\end{eqnarray}}
\def\nn{\nonumber}
\def\half{{1\over 2}}
\def\rhalf{{1\over \sqrt 2}}
\def\calo{{\cal O}}
\def\call{{\cal L}}
\def\calm{{\cal M}}
\def\del{\delta}
\def\eps{\epsilon}
\def\lam{\lambda}

\def\anti{\overline}
\def\delfac{\sqrt{{2(\del-1)\over 3(\del+2)}}}
\def\heff{h'}
\def\square{\boxxit{0.4pt}{\fillboxx{7pt}{7pt}}\hspace*{1pt}}
    \def\boxxit#1#2{\vbox{\hrule height #1 \hbox {\vrule width #1
             \vbox{#2}\vrule width #1 }\hrule height #1 } }
    \def\fillboxx#1#2{\hbox to #1{\vbox to #2{\vfil}\hfil}   }

\def\braket#1#2{\langle #1| #2\rangle}
\def\gev{~{\rm GeV}}
\def\gam{\gamma}
\def\sn{s_{\vec n}}
\def\sm{s_{\vec m}}
\def\mm{m_{\vec m}}
\def\mn{m_{\vec n}}
\def\mh{m_h}
\def\sumn{\sum_{\vec n>0}}
\def\summ{\sum_{\vec m>0}}
\def\vl{\vec l}
\def\vk{\vec k}
\def\ml{m_{\vl}}
\def\mk{m_{\vk}}
\def\gp{g'}
\def\gt{\tilde g}
\def\hw{{\hat W}}
\def\hz{{\hat Z}}
\def\ha{{\hat A}}

\def\yy{{\cal Y}_\mu}
\def\yyt{{\tilde{\cal Y}}_\mu}
\def\lq{\left [}
\def\rq{\right ]}
\def\dmu{\partial_\mu}
\def\dnu{\partial_\nu}
\def\dmus{\partial^\mu}
\def\dnus{\partial^\nu}
\def\gp{g'}
\def\gpt{{\tilde g'}}
\def\gs{g''}
\def\ggs{\frac{g}{\gs}}
\def\eps{{\epsilon}}
\def\tr{{\rm {tr}}}
\def\V{{\bf{V}}}
\def\W{{\bf{W}}}
\def\Wt{\tilde{ {W}}}
\def\Y{{\bf{Y}}}
\def\Yt{\tilde{ {Y}}}
\def\L{{\cal L}}
\def\s{s_\theta}
\def\st{s_{\tilde\theta}}
\def\c{c_\theta}
\def\ct{c_{\tilde\theta}}
\def\gt{\tilde g}
\def\et{\tilde e}
\def\At{\tilde A}
\def\Zt{\tilde Z}
\def\Wpt{\tilde W^+}
\def\Wmt{\tilde W^-}

\section{Introduction \label{sec:0}}

Higgsless models
\cite{Csaki:2003dt,Agashe:2003zs,Csaki:2003zu,Nomura:2003du}\cite{Barbieri:2003pr,Burdman:2003ya,
Cacciapaglia:2004jz,Davoudiasl:2004pw,Barbieri:2004qk}
have been recently  considered
as an alternative
to the standard electroweak symmetry breaking mechanism.
The corresponding  effective  theories  are strongly interacting
and share some  similarities with the  previously proposed  technicolor models.
Higgsless models are formulated as gauge theories in a five dimensional
space and, after decompactification, describe a tower of Kaluza Klein
(KK) excitations
of the standard electroweak gauge bosons. These theories can also be
understood as four dimensional deconstructed
\cite{Arkani-Hamed:2001ca,Arkani-Hamed:2001nc,Hill:2000mu,Cheng:2001vd}
\cite{Son:2003et,
Foadi:2003xa,Hirn:2004ze,Casalbuoni:2004id,Chivukula:2004pk,Chivukula:2004af,
Georgi:2004iy,Perelstein:2004sc}
theories in the context of linear moose models.
One of the interesting features of the Higgsless models is the
possibility to delay  the unitarity violation scale
via the exchange of massive KK modes \cite{Csaki:2003dt,SekharChivukula:2001hz,
Chivukula:2002ej,Chivukula:2003kq,DeCurtis:2002nd,DeCurtis:2003zt,
Abe:2003vg,Papucci:2004ip}.
However, in   the simplest version of these models,  it is difficult
to reconcile a delayed unitarity with the electroweak constraints:
in fact the $\eps_3$ parameter
 tends to get a large contribution. For instance in
 the framework of    models with only ordinary fermions it is possible to get small or
zero $\eps_3$ \cite{Casalbuoni:2004id}, at the expenses of having
a unitarity bound as in the Standard Model (SM)
without the Higgs, that is of the
order of 1 $TeV$. A recent solution to
the $\eps_3$ problem which does not spoil the unitarity
requirement at low scales, has been found by delocalizing the 
fermions in five dimensional theories
\cite{Cacciapaglia:2004rb,Foadi:2004ps}. In this paper we consider
a linear moose model and we try to obtain a solution by
introducing direct couplings (allowed by the symmetry of the
model) between ordinary left-handed fermions and the gauge vector
bosons  along the moose string. This is possible by defining these
couplings in terms of a product of  non linear $\sigma$-model scalar fields
which, in the continuum limit becomes a Wilson line. These
interactions have been previously considered within a simple
version of the moose models, the so-called BESS model
\cite{Casalbuoni:1985kq,Casalbuoni:1986vq}. Since the contribution
to the $\eps_3$ parameter coming from fermions can be of opposite
sign with respect to the one coming from the heavy vector mesons,
typical of the moose, in principle there can be cancellations,
though at the expenses of some fine tuning. This implies that the
masses of the heavy vector mesons can be kept sufficiently low  such that one can raise up the unitarity limit. In
our solution the fermions live in four dimensions and no fermion KK
excitations are present.

After reviewing the linear moose framework in Section II we introduce the new
couplings of the left-handed fermions to the gauge bosons in Section III.
In Section IV we
study the low energy limit of the model and we derive the
corresponding effective lagrangian containing only the SM fields.
The kinetic terms of the effective lagrangians are not in the
canonical form, therefore in Section V we proceed to a finite
renormalization of the fields. In Section VI we calculate the $\eps$
parameters in terms of the coupling constants appearing in the original
model. In Section VII we study some particular models according to
the variation of the couplings along the string, and we show that
there is in fact some space for cancellation, satisfying the experimental bounds. In Section VIII we study the
continuum limit. Conclusions are given in Section IX. In Appendix A we decouple the heavy particles by
finding explicit expressions for the corresponding fields in terms
of the SM gauge fields. Finally in Appendix B we derive the form of the
 low-energy
four-fermion interaction coming from
 the direct couplings of the fermions to the gauge bosons.
This term  provides a contribution to the
definition of the Fermi constant.

\section{Review of the  linear moose model for the electroweak symmetry breaking}
\label{linear}

Let us briefly review the linear moose model based on the $SU(2)$ symmetry.
Following   the idea of  dimensional deconstruction
\cite{Arkani-Hamed:2001ca,Arkani-Hamed:2001nc,Hill:2000mu,Cheng:2001vd},
  the hidden gauge symmetry
 approach applied  to the
strong interactions
\cite{Bando:1985ej,Bando:1988ym,Bando:1988br,Son:2003et,Hirn:2004ze}
 and
to the electroweak symmetry breaking
\cite{Casalbuoni:1985kq,Casalbuoni:1989xm,Hirn:2004ze},  we consider
$K+1$ non linear $\sigma$-model scalar fields $\Sigma_i$, ${i=1,\cdots ,K+1}$,
$K$ gauge groups, $G_i$, ${i=1,\cdots ,K}$
 and a global symmetry $G_L\otimes
G_R$. A minimal model of electroweak symmetry breaking
 is obtained by choosing $G_i=SU(2)$,
$G_L\otimes G_R=SU(2)_L\otimes SU(2)_R$. The SM
gauge group
  $SU(2)_L\times U(1)_Y$ is obtained by gauging
a subgroup of $G_L\otimes G_R$. The $\Sigma_i$ fields can be
parameterized as $\Sigma_i=\exp{(i/(2f_i)\vec \pi_i\cdot \vec
\tau})$ where $\vec \tau$ are the Pauli matrices and $f_i$ are
$K+1$ constants that we will call link couplings.

The transformation properties of the $\Sigma_i$ fields are the following
\bea
&&\Sigma_1\to L\Sigma_1 U_1^\dagger,\nn\\
&&\Sigma_i\to U_{i-1}\Sigma_i U_i^\dagger\
,\,\,\,\,\,\,\,i=2,\cdots,K,
\nn\\
&&\Sigma_{K+1}\to U_K\Sigma_{K+1} R^\dagger,
\label{transfs}
\eea
with $U_i\in
G_i$, $i=1,\cdots,K$, $L\in G_L$, $R\in G_R$.

The lagrangian of the linear moose model  for the gauge fields
is given by \be {\cal L}=\sum_{i=1}^{K+1}f_i^2{\rm
Tr}[D_\mu\Sigma_i^\dagger D^\mu\Sigma_i]-\frac 1 2\sum_{i=1}^K{\rm
Tr}[(F_{\mu\nu}^i)^2] -\frac 1 2{\rm Tr}[(F_{\mu\nu}(\Wt))^2 -\frac
1 2{\rm Tr}[(F_{\mu\nu}(\Yt))^2] , \label{lagrangian:l} \ee
with
the covariant derivatives  defined as follows \bea
&D_\mu\Sigma_1=\de_\mu\Sigma_1-i\gt \Wt_\mu\Sigma_1+i\Sigma_1 g_1
V_\mu^1,&\nn\\
&D_\mu\Sigma_i=\de_\mu\Sigma_i-ig_{i-1}V_\mu^{i-1}\Sigma_i+i\Sigma_i
g_i V_\mu^i,&\,\,\,\,\,\,\,i=2,\cdots,K,\nn\\
&D_\mu\Sigma_{K+1}=\de_\mu\Sigma_{K+1}-ig_{K}V_\mu^{K}\Sigma_{K+1}+i
\gpt\Sigma_{K+1}\Yt_\mu,&
\label{covderivative}
\eea
where $V_\mu^i=V_\mu^{ia}\tau^a/2$ and $g_i$ are the gauge fields and gauge coupling
constants associated to the groups $G_i$, $i=1,\cdots ,K$,
and $\Wt_\mu=\Wt_\mu^{a}\tau^a/2$, $\Yt_\mu=\yyt\tau^3/2$ are the gauge fields associated
to $SU(2)_L$ and $U(1)_Y$ respectively.

\begin{figure}[h] \centerline{
\epsfxsize=12cm\epsfbox{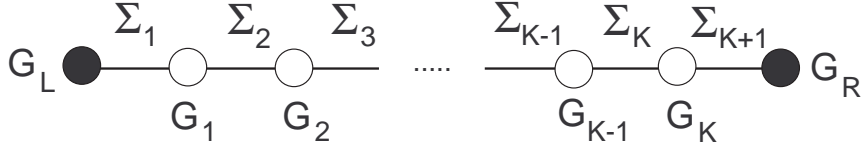} } \caption {{\it  The linear
moose model. \label{fig:1} }}
\end{figure}

The model described by the
lagrangian given in eq.(\ref{lagrangian:l})
is represented in Fig. \ref{fig:1}. Notice that the
field defined as
\be U=\Sigma_1\Sigma_2\cdots\Sigma_{K+1}
\label{chiral}\ee
is the usual chiral field: in fact
it transforms as
$
U\rightarrow LUR^\dagger
$
and it is invariant under the $G_i$ transformations.

The mass matrix of the gauge fields can be obtained by choosing
$\Sigma_i=I$ in eq.(\ref{lagrangian:l}). We find
\be {\cal
L}_{\rm mass}=\sum_{i=1}^{K+1}f_i^2{\rm Tr}[(g_{i-1}V_\mu^{i-1}-g_i
V_\mu^i)^2]\equiv \frac 1 2\sum_{i,j=0}^{K+1}(M_2)_{ij}V_\mu^i
V^{\mu j}, \label{lmass}\ee 
where we have defined
$V_\mu^0= \Wt_\mu$, $V_\mu^{K+1}= \Yt_\mu$, $g_0=\gt$, $g_{K+1}=\gpt$,
$f_0=f_{K+2}
=0$ and
\be
(M_2)_{ij}=g_i^2(f_i^2+f_{i+1}^2)\delta_{i,j}-g_i g_{i+1}f_{i+1}^2
\delta_{i,j-1}-g_j g_{j+1}f_{j+1}^2  \delta_{i,j+1}. \label{m2}\ee

\section{Couplings to fermions}

In the following we will consider standard model fermions, that is:
 left-handed fermions  $\psi_L$ as  $SU(2)_L$
doublets and  singlet right-handed fermions 
$\psi_R$. The standard couplings are given by 
\be {\cal
L}_{fermions}=\bar\psi_Li\gamma^\mu (\de_\mu+i \gt\Wt_\mu+ \frac i
2 \gpt (B-L) \yyt)\psi_L+ \bar\psi_Ri\gamma^\mu (\de_\mu+i \gpt
\Yt_\mu+ \frac i 2 \gpt(B-L) \yyt)\psi_R 
\label{lfermion}\ee 
where B(L) is the
baryon (lepton) number. These are the only fermions introduced in
this model and they are coupled to the SM gauge fields through the
groups $SU(2)_L$ and $U(1)_Y$ at the ends of the chain.

We can introduce also direct couplings of the $\psi_L$ fermions to the
field $V_\mu^i$  by generalizing the procedure of
\cite{Casalbuoni:1985kq,Casalbuoni:1986vq}. For each $\psi_L$ we can
construct the following $SU(2)$ doublets
 \be \chi^i_L=\Sigma_i^\dagger
\Sigma_{i-1}^\dagger\cdots \Sigma_1^\dagger
\psi_L\,,\,\,\,i=1,\dots,K\,.
\label{eq:8} \ee 
These fields transform
under eqs.(\ref{transfs}) as \be \chi^i_L\rightarrow U_i\Sigma_i^\dagger
U_{i-1}^\dagger U_{i-1}\Sigma_{i-1}^\dagger U_{i-2}^\dagger\cdots
U_1^\dagger U_1 \Sigma_1^\dagger L^\dagger \psi_L'=
U_i\Sigma_i^\dagger \cdots \Sigma_1^\dagger L^\dagger L \psi_L=U_i
\chi_L^i \,.
\ee Therefore we can add  
to the fermion lagrangian in eq.(\ref{lfermion}) 
the
following term containing direct left-handed fermion couplings to $V^i_\mu$ 
which is invariant under the symmetry transformation of the model:
\be \sum_{i=1}^K b_i
\bar\chi_L^i i\gamma^\mu (\de_\mu+i g_i V_\mu^i+ \frac i 2\gpt (B-L)
\yyt)\chi_L^i \ee where $b_i$ are $K$ dimensionless parameters. In
the unitary gauge $\Sigma_i=I$ and therefore the total fermion
lagrangian is given by \bea {\cal
L}_{fermions}^{tot}&=&\bar\psi_Li\gamma^\mu (\de_\mu+i \gt\Wt_\mu+
\frac i 2
\gpt(B-L) \yyt)\psi_L\nn\\
&+&
\sum_{i=1}^K b_i \bar\psi_L i\gamma^\mu (\de_\mu+i g_i V_\mu^i+ \frac i 2
\gpt(B-L) \yyt)\psi_L\nn\\
&+&\bar\psi_Ri\gamma^\mu (\de_\mu+i \gpt  \Yt_\mu+ \frac i 2
\gpt(B-L) \yyt)\psi_R\,.
\eea
The canonical kinetic term for fermions is obtained
by the following redefinition
\be
\psi_L\rightarrow \frac 1{\sqrt{1+\sum^K_{i=1} b_i}}\psi_L\,,
\ee
so that the final fermion coupling lagrangian is given by
\bea
{\cal L}_{fermions}^{tot}&=&\bar\psi_L i\gamma^\mu \de_\mu \psi_L +
\bar\psi_R i\gamma^\mu \de_\mu \psi_R\nn\\
&+&\frac 1 {{1+\sum^K_{i=1} b_i}}\bar\psi_L i\gamma^\mu
(i \gt\Wt_\mu+ \frac i 2 \gpt
(B-L) \yyt)\psi_L\nn\\
&+&
\sum_{i=1}^K  \frac {b_i}{{1+\sum^K_{j=1} b_j}}\bar\psi_L i\gamma^\mu
(i g_i V_\mu^i+ \frac i 2 \gpt
(B-L) \yyt)\psi_L\nn\\
&+&\bar\psi_Ri\gamma^\mu (i \gpt \Yt_\mu+ \frac i 2
\gpt(B-L) \yyt)\psi_R\label{eq:13}\,.
\eea



\section{The low-energy limit}

Let us study the effects of the $V_i$ ($i=1,\dots,K$) particles in the low-energy limit.
This can be done by eliminating the $V_i$ fields with the solution
of their equations of motion for  $g_i\gg 1$, limit that corresponds
to heavy masses for the  $V_i$ fields (see eq.(\ref{m2})).
 In fact in this
limit the kinetic term of the new resonances is negligible.
The corresponding 
effective theory will be  considered   up to order $(1/g_i)^2$.

Let us solve the equations of motion for the field $V_i$ in terms of $\Wt$ and $\Yt$
(see Appendix A). For the moment being we
 neglect fermion current contributions which give current-current
interactions in the effective lagrangian, these will be considered later on. By separating
charged and neutral components, we get
\be V_i^\alpha=\frac 1
{g_i}(\gt\Wt^\alpha z_i) \,,~~~ \alpha=1,2\,,\ee
\be V_i^3=\frac 1 {g_i}(\gpt \yyt y_i+\gt\Wt^3z_i)\,,
\ee
where we have, for convenience, introduced the following variables
\be z_i=\sum^{K+1}_{j=i+1}x_j,~~~x_i=\frac{f^2}{f_i^2},~~~\frac
1{f^2}=\sum_{i=1}^{K+1}\frac 1{f_i^2},~~~\sum_{i=1}^{K+1}x_i=1,~~~
y_i=1-z_i\,.\ee
By using the standard linear combinations
\bea
{\At}_\mu &=&  \st {\Wt}^3_\mu  +\ct \yyt\,,\nn\\
{\Zt}_\mu &=&   \ct {\Wt}^3_\mu -\st  \yyt\,,
\eea
with $\st$ and $\ct$ defined as in the SM,
\be
\et=\gt\st=\gpt\ct
\ee
and by substituting in the quadratic part of the kinetic
 lagrangian, we obtain  
\bea
\L^{kin~(2)}_{eff}({\tilde W}^\pm,\At,\Zt) &=&
 -\f{1}{4} (1+z_\gamma)\At_{\mu\nu}\At^{\mu\nu}
-\f{1}{2} (1+z_w){\tilde W}_{\mu\nu}^+ {\tilde W}^{\mu\nu-} \nn\\
&-& \f{1}{4} (1+z_z)\Zt_{\mu\nu}\Zt^{\mu\nu}+\f{1}{2} z_{z\gamma}
\At_{\mu\nu}\Zt^{\mu\nu} \label{kin}\,,
\eea
where $O_{\mu\nu}=\dmu O_\nu-\dnu
O_\mu$, ($O={\tilde W}^\pm,\At,\Zt$) and \be z_\gamma =
\sum_{i=1}^K\Big(\f{\et}{g_i}\Big)^2,~~~~~ z_w = \sum_{i=1}^K\Big(\f
{\gt}{g_i}\Big)^2z_i^2,~~~~~ z_z =
\f{\et^2}{\st^2\ct^2}\sum_{i=1}^K\f{1}{g_i^2}\Big(
z_i-\st^2\Big)^2,~~~~~ z_{z\gamma} =  -\f{\et^2}{\st\ct}\sum_{i=1}^K
\f{1}{g_i^2}\Big( z_i-\st^2\Big) \,.\ee


By making use of  the solutions of the equations of motion for $V_i$ in the fermion lagrangian,
we obtain
\bea
\L_{eff}^{charged} &=& -\frac{\et}{\sqrt{2} \st}
\big(1-\frac{b}{2}\big)\overline\psi_d
     \gamma^\mu\frac{1-\gamma_5}{2}\psi_u \Wmt_\mu +~h.c.\,,\label{eq:21}\\
\L_{eff}^{neutral} &=& -\frac{\et}{\st \ct} \big(1-\frac{b}{2}\big)\overline\psi
     \gamma^\mu\Big[ T^3_L \frac{1-\gamma_5}{2}-\frac{Q \st^2}{ \big(1-\frac{b}{2}\big)}\Big]
       \psi \Zt_\mu - \et \overline\psi \gamma^\mu Q \psi
       \At_\mu\,,\label{eq:22}
\eea
with
\be b=2 \frac {\sum^K_{i=1} b_i y_i}{1+\sum^K_{i=1} b_i}
\ee
and $T_L^3\psi_L=\tau_3/2 \psi_L$, $T_L^3\psi_R=0$.

As shown in Appendix B, the additional fermion direct couplings to $V_i$ give rise
to 
the following current-current interaction term:  
\be{\cal L}^{ quart}_{eff}
=\beta\sum_{a=1}^3\left(\bar\psi_L\gamma^\mu\frac{\tau^a}2\psi_L\right)^2\label{eq:25t}
\ee 
with 
\be \beta=\frac 1{8f^2}\left(\bar b_K-b\right)^2-\frac 1
{8f^2}\sum_{i=1}^K x_{i+1}\bar b_i^2\ee
and
\be
\bar
b_i=2\frac{\sum_{j=1}^i b_j}{1+\sum_{j=1}^K b_j}~~~(i=1,\cdots , K)\,.\ee

\section{Fields and couplings renormalization}

The corrections to the quadratic part of the kinetic lagrangian
given in eq.(\ref{kin})  are $U(1)_{em}$ invariant and produce a
wave-function renormalization of $\At_\mu,\Zt_\mu,{\tilde
W}_\mu^\pm$ plus a mixing term $\At_\mu-\Zt_\mu$. Notice that in
general there could be two other renormalization terms: $\delta
M_W^2 \Wpt_\mu{\tilde W}^{\mu-}$ and  $\delta M_Z^2 \Zt_\mu\Zt^\mu$
which, however, are zero in this model.
To identify the physical quantities we define new fields in such a way
to have canonical kinetic terms and to cancel the mixing term
$\At_\mu-\Zt_\mu$. They are given by the following relations:
\bea
\At_\mu &=& (1-\frac{z_\gamma}{2}) A_\mu+z_{z_\gamma} Z_\mu\,,\nn\\
{\tilde W}^\pm_\mu &=& (1-\frac{z_w}{2}) W_\mu^\pm\,,\nn\\
\Zt_\mu &=& (1-\frac{z_z}{2}) Z_\mu\,.\label{eq:27} \eea

Let us study the effects of this renormalization.
First of all for the
mass terms we get:
\be
- f^2\tr(\Wt_\mu-\Yt_\mu)^2=-{\tilde M}_W^2 (1-z_w) W^{\mu+}W^-_\mu
-\frac{1}{2}{\tilde M}_Z^2 (1-z_z) Z^\mu Z_\mu
\ee
where, for comparison with  the SM results, we have defined  $f=v/2$ and
\be
{\tilde M}_W^2 =\frac{v^2}{4}\gt^2\,, ~~~~~~~~
{\tilde M}_Z^2={\tilde M}_W^2/\ct^2\,.
\ee

Also, the field renormalization affects all the couplings of the
standard gauge bosons to the fermions. By substituting eq.(\ref{eq:27}) in eqs.(\ref{eq:21}) and (\ref{eq:22}) we get \bea
\L_{eff}^{charged} &=& -\frac{\et}{\sqrt{2} \st}
\big(1-\frac{b}{2})(1-\frac{z_w}{2}\big)
     \overline\psi_d
     \gamma^\mu\frac{1-\gamma_5}{2}\psi_u W^-_\mu +~h.c.\label{eq:30}\,,\\
\L_{eff}^{neutral} &=& -\frac{\et}{\st \ct} \big(1-\frac{b}{2})(1-\frac{z_z}{2}\big)
      \overline\psi
     \gamma^\mu\Big[ T^3_L \frac{1-\gamma_5}{2}-Q \st^2
\frac{\dd 1
      -\frac{\ct}{\st}z_{z \gamma}}{\dd 1-\frac{b}{2}}\Big]
       \psi Z_\mu\nn\\
     & &  - \et \big(1-\frac{z_\gamma}{2}\big)
        \overline\psi \gamma^\mu Q \psi A_\mu\label{eq:31}\,.
\eea

We see that the physical constants as the electric charge, the Fermi constant
and the mass of the $Z$  must be redefined in terms of the parameters appearing
in our effective lagrangian. They are identified as follows
\bea
e &=& \et \big(1-\frac{z_\gamma}{2}\big)\,,\nn\\
M_Z^2 &=& {\tilde M}_Z^2 (1-z_z)\,.
\eea
Concerning the Fermi constant $G_F$, it is evaluated from the $\mu$-decay
process. Taking into account the modified
charged current coupling,
the $W$ mass
\be
M_W^2 = {\tilde M}_W^2 (1-z_w)
\ee
and the charged current-current interaction
 we obtain
\bea \frac{G_F}{\sqrt{2}}&=&\f 1 8 \gt^2 \Big (1-\f b 2\Big)^2\frac
{1-z_w}{M_W^2} +\f 1 4 \beta 
\label{GF}\eea where we have used eqs.
(\ref{eq:30}) and (\ref{eq:25t}).

Following \cite{Altarelli:1993sz} we define $\s$ by
\be
\frac{G_F}{\sqrt{2}}=\f {e^2}{8\s^2\c^2 M^2_Z}\,.
\ee
By comparing with eq.(\ref{GF}) we
get  
\be s_{2\tilde\theta}=s_{2\theta}\sqrt{X}
\ee 
where
\begin{equation}
X = \frac{\left (1-\f {\dd b} {\dd 2}\right )^2}{1-\f {\dd \sqrt{2}\beta}{\dd 4G_F}} (1 +z_\gamma-z_z)\,.
\end{equation}
More explicitly
\be 
\st^2 =\frac{1}{2} - \frac{1}{2}\sqrt{1 -\f{4\pi\alpha}{\sqrt{2}G_FM_Z^2}X}
\label{eq:39}
\ee
with $\alpha$  the fine structure constant. Notice that for $X=1$ we recover
the  standard definition for $\s$.

\section{The $\eps$ parameters}

The corrections to the tree-level SM results are usually parameterized in terms of
a set of parameters called $\eps_{1,2,3}$  that can be
obtained from
$\Delta r_W$, $\Delta\rho$ and $\Delta k$ \cite{Altarelli:1993sz,Altarelli:1998et}.
Let us start from $\Delta r_W$ defined by:
\be
\frac{M^2_W}{M^2_Z}=
 \c^2\left[1-\frac{\s^2}{c_{2\theta}}\Delta r_W\right]\,.
\ee
{}From the relation ${\tilde M}^2_W=
{\tilde M}^2_Z\ct^2$, and using eq.(\ref{eq:39}), we get
\be
\frac{M^2_W}{M^2_Z}=
 \c^2\left[\left(1+ z_z-z_w\right)\left(\frac{1}{2} +\frac{1}{2}\sqrt{1 - \f{4\pi\alpha}
 {\sqrt{2}G_FM_Z^2} X}\right)\right]
\ee so, for comparison, 
\bea \Delta
r_W&=&
\frac{c_{2\theta}}{\s^2}\Big[ 1-\f {\ct^2} {\c^2}(1+z_z-z_w)\Big]\nn\\
&=&\frac{c_{2\theta}}{\s^2\c^2}\left[\c^2-\left(\frac{1}{2}
+\frac{1}{2}\sqrt{1 - \f{4\pi\alpha}{\sqrt{2}G_FM_Z^2} X}\right)(1+
z_z-z_w) \right]
\label{deltarW}
\eea

The definitions of $\Delta\rho$ and $\Delta k$
are given in terms of the neutral current
couplings to the $Z$ gauge boson
\be
\L^{neutral}(Z)=-\frac{e}{\s\c}\Big(1+\frac{\Delta\rho}{2}\Big)Z_\mu\overline\psi
[\gamma^\mu g_V+\gamma^\mu \gamma_5g_A]\psi
\ee
with
\bea
g_V &=& \frac{T^3_L}{2}-s^2_{\bar\theta} Q\,,\nn\\
g_A &=& -\frac{T^3_L}{2}\,,\nn\\
s^2_{\bar\theta} &=& (1+\Delta k) \s^2\,.
\eea

For comparison with eq.(\ref{eq:31}) we obtain

\bea
\Delta\rho &=& 2 \left[ \frac{\s \c}{\st \ct} \left(1 - \frac{b}{2} \right)\left(1 - \frac{z_z}{2} \right)\left(1 - \frac{z_\gamma}{2} \right)^{-1}  - 1\right] = 2 \left( \sqrt{1 - \frac{\beta}{2 \sqrt{2}}\frac{1}{G_F}} - 1 \right)\,,  \nn\\
\Delta k &=& -1 + \frac{\st^2}{\s^2}\frac{1}{\left( 1 - \frac{b}{2}\right) }\left( 1 - \frac{\ct}{\st}{z_{z \gamma}}\right)\,,
\label{deltarhok}
\eea
with $\st$ given in eq.(\ref{eq:39}).
Finally we can compute the new physics contribution to the
 $\eps$ parameters \cite{Altarelli:1993sz}
\bea
\eps_1^N &=& \Delta\rho\,,\nn\\
\eps_2^N &=& \c^2\Delta\rho+\f{\s^2}{c_{2\theta}}\Delta r_W-2 \s^2\Delta k\,,\nn\\
\eps_3^N &=& \c^2\Delta\rho+c_{2\theta}\Delta k \,.
\label{epsdef}
\eea
Expanding up to the first order in $b_i$ and neglecting
 terms $O(b_i/g_i^2))$, we get
\bea
\eps_1^N &\simeq& 0\,,\nn\\
\eps_2^N &\simeq& 0\,,\nn\\
\eps_3^N &\simeq&\sum_{i=1}^{K}y_i(\f {e^2 }{\s^2 g_i^2} z_i-b_i) \,.
\label{epsappr}\eea
This final expression suggests that the introduction of
the $b_i$ direct fermion couplings to $V_i$ can compensate for the
contribution of the tower of gauge vectors to $\eps_3$. This would
reconcile  the Higgsless model with the electroweak precision
measurements by fine-tuning the direct fermion couplings.

\section{Particular models}

In this section we consider the bounds on the parameter space of the linear moose model in some examples, by comparing the theoretical predicted values for 
the $\eps$ parameters with 
the experimental values
\cite{Barbieri:2004qk}
\bea
\eps_1 &=& (5\pm 1.1 )\times 10^{-3}\,,\nn\\
\eps_2 &=& (-8.8\pm 1.2 )\times 10^{-3}\,,\nn\\
\eps_3 &=& (4.8\pm 1.0)\times 10^{-3}\,. \eea
We assume  the same radiative corrections for the
$\eps$ parameters as in the SM
with a cutoff $m_H=1$ TeV and $m_{top}=178$ GeV.
We used the exact formulas in $b_i$ given in eqs.(\ref{epsdef}),
(\ref{deltarW}) and (\ref{deltarhok}).

\begin{figure}[h] \centerline{
\epsfxsize=7cm\epsfbox{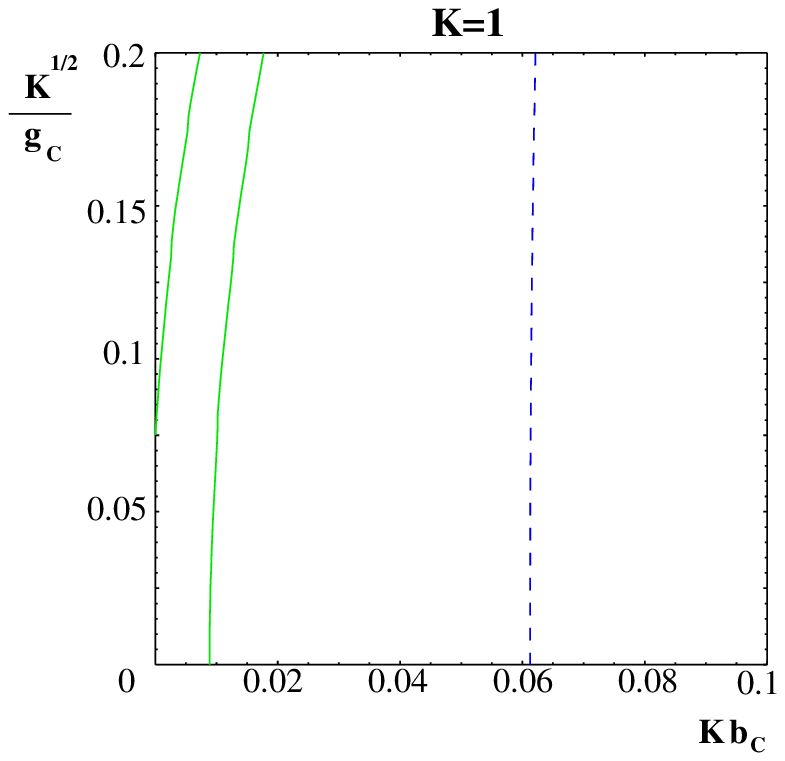} 
\epsfxsize=7cm\epsfbox{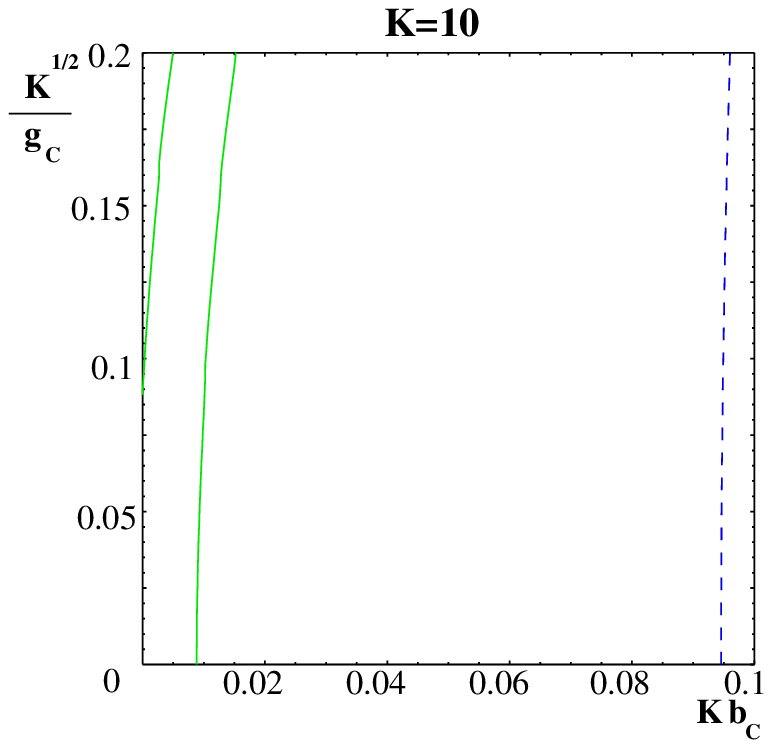}
} \caption {{\it
 95\% CL bounds on the parameter space ($Kb_c,\sqrt{K}/g_c$) from the experimental values
of $\eps_2$ and $\eps_3$ for  $K=1$ (left) $K=10$ (right), assuming the SM
radiative corrections for $m_H=1$ TeV
and $m_{top}=178$ GeV.
The allowed parameter space from $\eps_2$ is the region to the left of the
 dashed line and from $\eps_3$
the region between the  continuous  lines.
\label{fig:2} }}
\end{figure}
We first study the simplest model with all $f_i=const=f_c$, $g_i=const=g_c$
and $b_i=const=b_c$.

 For $K=1$ and $K=10$ gauge
groups we get the bounds from $\epsilon_2$ and $\eps_3$
 shown in Fig. \ref{fig:2}. The experimental value for $\eps_1$ gives a $95\%$ CL
bound independent on $g_c$, $b_c\lesssim 0.14$ for $K=1$ and $b_c\lesssim 0.025$
for $K=10$.

The most stringent bound comes from $\eps_3$, and it shows that  a
region exists where it is possible to satisfy the electroweak
constraints for small values of $Kb_c$. 
As shown in the Fig. \ref{fig:2}, the limits  in these variables
do not strongly depend on $K$ because of a scaling property. 
 It is obvious that for increasing $K$ the allowed region in the
parameter space $(b_c,1/g_c)$
shrinks.

\begin{figure}[h] \centerline{
\epsfxsize=7cm\epsfbox{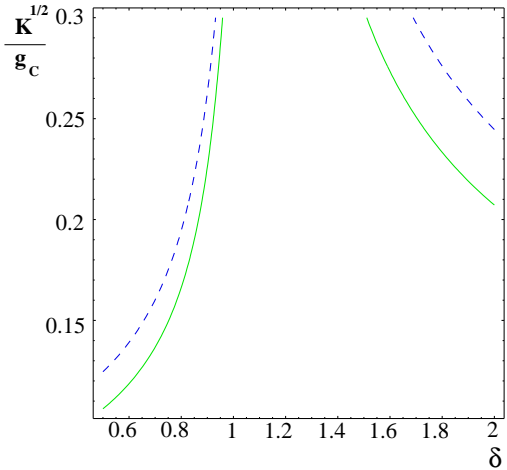} } \caption {{\it
 95\% CL bounds on the parameter space ($\delta,\sqrt{K}/g_c$) from the experimental value
of $\eps_3$ for $K=1$ (continuous line), $K=10$
(dash line), assuming the radiative corrections for $m_H=1$ TeV
and $m_{top}=178$ GeV. The allowed parameter space is the region between the corresponding
lines.\label{fig:4} }}
\end{figure}

The approximate expression for  $\eps_3$ given in eq.(\ref{epsappr})
suggests also the following choice for $b_i$,
\be
b_i=\delta\f {e^2 }{\s^2 g_i^2} z_i=\delta\f {e^2 }{\s^2 g_i^2} (1-y_i)\label{eq:50}
\ee
which
gives $\eps_3^N\simeq 0$ for $\delta=1$ for small $b_i$.
Assuming again  $f_i=f_c$, $g_i=g_c$, this means
\be
b_i=\delta\f {e^2}{\s^2g_c^2}\left(1-\frac {i}{K+1}\right )\,.
\ee
In Fig. \ref{fig:4}  we plot the
 95\% CL bounds on the parameter space ($\delta,\sqrt{K}/g_c$) from the experimental value
of $\eps_3$ for $K=1$ (continuous line), $K=10$
 (dash line), assuming the radiative corrections for $m_H=1$ TeV
and $m_{top}=178$ GeV. For $K\gg 10$ the allowed region is nearly independent on $K$. Very loose bounds are obtained from $\eps_1$ and $\eps_2$.

In conclusion, by fine tuning every direct fermion
coupling in each site in a way to compensate the corresponding contribution
to $\eps_3$ from the linear moose, a sizeable region in the 
parameter space is left.

\section{Continuum limit}

It is known that the discretization of a gauge theory lagrangian
 in a 4+1 dimensional space-time
along the fifth dimension (the segment of length $\pi R$)
 gives rise to  a linear moose chiral lagrangian after a suitable
identification of the gauge and link couplings
\cite{Arkani-Hamed:2001nc,Arkani-Hamed:2001ca,Hill:2000mu,Cheng:2001vd,
Randall:2002qr}.
We consider the continuum limit $a\to 0$, $K\to\infty$
 with the condition
$K a=\pi R$, where $\pi R$ is the length of the segment in the fifth
dimension. We would like to discuss what is the continuum limit for
the direct fermionic couplings when we choose the $b_i$'s according
to the eq.(\ref{eq:50}) with $\delta=1$ for simplicity. By
defining 
\be\lim_{a\to 0} \frac {b_i} a=b(y),~~~\lim_{a\to 0}
a{f_i^2}=f^2(y),~~~\lim_{a\to 0}a g_i^2=g_5^2(y)
\ee 
we find,
assuming $g_5(y)=g_5$ with $g_5$ a constant \be b(y)=
\frac{e^2 }{\s^2g_5^2}\int_y^{\pi R} dt\frac{f^2}{f^2(t)}\label{eq:53}\ee
 with
\be \frac 1{f^2}=\int_0^{\pi R} \frac{dy}{f^2(y)}\,.\ee 
From eq.
(\ref{eq:53}) we see that \be b(0)=\frac{e^2 }{\s^2 g_5^2},~~~~b(\pi R)=0\,.\ee 
Therefore the direct fermionic
coupling decreases along the fifth dimension going from the brane
located at $y=0$ to the brane at $y=\pi R$. For the case of constant
$f(y)=\bar f$ we find 
\be b(y)=\frac{e^2 }{\s^2 g_5^2}\left(1-\frac
y{\pi R}\right)\,.\ee 
With this choice the contribution from the new delocalized 
fermion interactions to
$\eps_3^N$ is given by
\be
\eps_3^N\vert_{ferm}=-\frac 1 {\pi R}\int_0^{\pi R}dy ~y b(y)=-\frac {e^2 }
{\s^2 g_5^2}\frac {\pi R}{6}
\ee
which is just the opposite of the contribution to $\eps_3^N$
in the linear moose \cite{Foadi:2003xa,Casalbuoni:2004id}.

Another interesting case corresponds to a
Randall-Sundrum metric along the fifth dimension \cite{Randall:1999vf,Randall:2002qr}, that is
\be
f(y)=\bar f e^{ky}\ee 
and we find 
\be b(y)=\frac{e^2 }
{\s^2 g_5^2}\frac{e^{-2\pi k R}-e^{-2ky}}{e^{-2\pi k R}-1}\,.\ee
In this case we get
\be
\eps_3^N\vert_{ferm}=-\int_0^{\pi R}dy \frac {e^{-2ky}-1}{e^{-2k \pi R}-1}  b(y)=
-\frac{e^2 }{\s^2 g_5^2}\frac 1 {4k}\frac{e^{4 k\pi R}-4 k\pi R e^{2 k\pi R}-1}
{(1-e^{2k\pi R})^2}
\ee
which is the opposite of the contribution from the gauge bosons
derived in \cite{Casalbuoni:2004id}. 

Summarizing,
in  the continuum limit,   left- and right-handed fermions live at the
opposite ends of the extra-dimension; they feel only the SM gauge
transformations. However, in the discrete, we have introduced an
interaction term invariant under all the symmetries of the model
which delocalizes the left-handed fermions in the continuum limit.
In fact, we have seen in eq.(\ref{eq:8}) that the fermionic fields
along the string are defined in terms of the operator \be
\Sigma_1\Sigma_2\cdots\Sigma_i\label{eq:59}\,.
\ee In five-dimensions
the fields $\Sigma$'s can be interpreted as the link variables along
the fifth dimension. As such they can be written in terms  the fifth
component of the heavy gauge fields $V$. As a consequence
 the operator given in eq.(\ref{eq:59}) becomes a Wilson
line 
\be \Sigma_1\Sigma_2\cdots\Sigma_i\to
P\left(\exp\left(i\int_0^y dt V_5(t,x)\right)\right)\,.\ee In this way
the original fermionic fields acquire a non-local interaction
induced by Wilson lines.

This non-local interaction is the origin of the possible negative
contribution to the parameter $\eps_3$.

\section{Conclusions}
\label{conclusions}

Models with replicas of gauge groups have been recently considered
because they  appear in the deconstruction of five dimensional gauge
models which have been used  to describe the electroweak breaking
without the Higgs
\cite{Arkani-Hamed:2001nc,Hill:2000mu,Cheng:2001vd,Son:2003et,
Foadi:2003xa,Hirn:2004ze,Casalbuoni:2004id,Chivukula:2004pk,Chivukula:2004af,
Georgi:2004iy,Perelstein:2004sc}.
The four dimensional description is based on the linear moose
lagrangians that were already proposed in technicolor and composite
Higgs models \cite{Georgi:1986hf}. In general these models satisfy
the constraints arising
 from the parameters $T$ and $U$ (or $\epsilon_1$ and $\epsilon_2$) due to
the presence of a custodial $SU(2)$ symmetry. However they
generally  give a correction of order ${\cal O}(1)$ (${\cal
O}(10^{-2}$)) to the parameter $S$ ($\epsilon_3$). In this paper
we have considered a linear moose based on replicas of $SU(2)$
gauge groups, with the electroweak gauge groups $SU(2)_L$ and
$U(1)_Y$ and ordinary fermions at the two ends of the moose string. 
Within this framework it seems that the
only way to satisfy the electroweak constraints is  the
one considered in \cite{Casalbuoni:2004id,Hirn:2004ze} which,  however,
sets   the unitarity violation scale around 1 TeV, as in the SM
without a Higgs. In this paper we have considered the possibility
of raising up this scale  by allowing the fermions to interact with
the gauge fields along the string. This is realized 
through the introduction of a string of scalar fields which, in
the continuum limit, is equivalent to delocalize the left-handed
fermions on the boundary with a Wilson line. This new non local
interaction gives a contribution to $\eps_3$ of opposite sign with
respect to the one coming from the gauge fields along the string.
Therefore, at the expenses of some fine tuning, it is possible to
satisfy the experimental limits. At the same time the scale of the
heavy vector bosons can be lowered allowing a corresponding increasing of the unitarity
bound.

After this work was completed, a related paper appeared
\cite{Chivukula:2005bn} where a partial delocalization of fermions
was  considered.

\appendix

\section{}
\label{appendixA}
The covariant derivatives of eq.(\ref{covderivative}) can be expressed in a
compact form, by defining $V_\mu^0= \Wt_\mu$,
$V_\mu^{K+1}=\tilde Y_\mu$, $ \tilde g =g_0$, $\gpt= g_{K+1}$,

\be
D_\mu\Sigma_i=\de_\mu\Sigma_i-ig_{i-1}V_\mu^{i-1}\Sigma_i+i\Sigma_i
g_i V_\mu^i,\,\,\,\,\,\,\,i=1,\cdots,K+1\,.
\ee
The lagrangian for the mass terms of the gauge fields
and their fermion couplings can be rewritten
 \be {\cal
L}_{mass}=\frac 1 2\sum_{i=1}^{K+1}f_i^2 B_{i-1}^2
-\sum_{i=0}^{K+1}J^a_i V_i^a .\label{A2}\ee The variables $B_i$
are the analogue in the discrete formulation of canonical momenta  and are
given by \be
 B_i= g_iV_i -g_{i+1}V_{i+1},~~~~i=0,\cdots,K,\ee
the $V_i^a=Tr(V_i \tau^a)$ and $J^a_i$ are the related fermionic
currents.
For $i=0,K+1$ $J_i$ are the SM fermionic currents, while
\be
J^a_i=- g_i \frac {b_i}{{1+\sum^K_{j=1} b_j}}\bar\psi_L \gamma^\mu \frac
{\tau^a} 2
\psi_L\,,\,\,\,\,\,\,i=1,\cdots,K\,.
\ee
Notice that the $K+1$ fields   $B_i$ are not independent, since
\be \sum_{i=0}^K B_i=
g_0V_0-g_{K+1}V_{K+1}
.\label{indep}\ee

 Therefore
we solve the equations of motion, which involve
three nearest neighborhoods, by solving first in the $B_i$'s and
then inverting the relation between the $B_i$'s and the fields
$V_i$. These equations involve only first neighborhoods. This is
the analogue of converting a second order differential equation in
a pair of first order equations.

The equations of motion can be written in the following form
\be
-f_i^2B_{i-1}+f_{i+1}^2B_i=L_i,~~~~i=1,\cdots,K,
\ee where we
have redefined the sources as \be L_i=\frac{J_i}{g_i}.\ee We can
solve for all the $B_i$, $i=1,\cdots,K$ in terms of $B_0$ finding
\be B_i=\frac 1{f_{i+1}^2}\left(\sum_{j=1}^i L_j+f_1^2
B_0\right),~~~~~i=1,\cdots,K.\label{eq:B7}\ee  It is convenient to
introduce the following variables \be \frac
1{f^2}=\sum_{i=1}^{K+1}\frac 1
{f_i^2},~~~~x_i=\frac{f^2}{f_i^2},~~~~i=1,\cdots,K+1,\ee \be
y_i=\sum_{j=1}^i x_j,~~~z_i=\sum_{j=i+1}^{K+1}x_j,\ee with  the
properties \be y_i+z_i=1,~~~~y_1=x_1,~~~~z_K=x_{K+1}.\ee

By summing the eqs.(\ref{eq:B7})
 over $i$ from 1 to $K$ and using eq.(\ref{indep}) we get a relation for
$B_0$ which can be easily solved obtaining \be
B_0=-\frac{x_1}{f^2}\sum_{i=1}^K
z_i\,L_i+x_1(g_0V_0-g_{K+1}V_{K+1}),\ee and \label{A11}\be
B_i=\frac{x_{i+1}}{f^2}\left(\sum_{j=1}^i\,y_j\,L_j-\sum_{j=i+1}^K\,z_j\,L_j
+f^2(g_0V_0-g_{K+1}V_{K+1}) \right).\label{A12}\ee By using  the
discrete step function given by 
\be
\theta_{i,j}=\Big\{^{\dd{1,~~~{\rm for}~ i\ge j}}_{\dd{0,~~~{\rm
for}~ i<j}}\ee
 we
can write 
\be
B_i=\frac{x_{i+1}}{f^2}\sum_{j=1}^K\left(\theta_{i,j}y_j-\theta_{j,i+1}z_j
\right)L_j +f^2(g_0V_0-g_{K+1}V_{K+1}),~~~i=0,\cdots,K\label{momenta}.
\ee 

Further we need to reexpress the
fields $V_i$ in terms of the $B_i$'s. We find
\bea
 V_i&=&\frac 1
{g_i}g_{K+1}V_{K+1}+
\frac 1
{g_i}\sum_{j=1}^K\theta_{j,i}B_j\,\nn\\
&=&\frac 1 {g_i}\Big[g_0V_0z_i+
g_{K+1}V_{K+1}y_i+\sum_{j=1}^K\theta_{j,i}x_{i+1}
\sum_{l=1}^K\left(\theta_{j,l}y_l-\theta_{l,j+1}z_l \right)L_l
\Big ] \,.\label{A15}\eea If we neglect fermion currents and we separate
charged and neutral components we obtain
\be V_i^\alpha=\frac 1
{g_i}(\gt\Wt^\alpha z_i) \,,\ee \be V_i^3=\frac 1 {g_i}(\gpt \tilde{\cal Y} y_i+\gt\Wt^3z_i)\,.
\ee

\section{}

In this appendix we evaluate the quartic terms in the fermion
fields arising after having eliminated the heavy fields $V_i$,
$i=1,\cdots, K$. Let us start again from the initial lagrangian
excluding the kinetic terms as given in eq.(\ref{A2}) 
\be {\cal
L}_{mass}=\frac 1 2\sum_{i=1}^{K+1}f_i^2 B_{i-1}^2
-\sum_{i=1}^{K}J^a_i V_i^a -J_0V_0-J_{K+1}V_{K+1}.
\ee The
solutions in eq.(\ref{momenta}) for the fields $B_i$ can be
expressed as 
\be B_i=\bar
B_i+x_{i+1}C,~~~C=(g_0V_0-g_{K+1}V_{K+1}),~~~i=0,\cdots,K.
\ee
where $\bar B_i$ are the solutions when the fields $V_0$ and
$V_{K+1}$ (the standard fields $\Wt$ and $\tilde Y$) are turned off, that
is 
\be \bar B_0=-\frac{x_1}{f^2}\sum_{i=1}^K z_i\,L_i,~~~ \bar
B_i=\frac{x_{i+1}}{f^2}\sum_{j=1}^K\left(\theta_{i,j}-z_j\right)\,L_j\,.
\ee
Let us notice that the terms in $J_0$, $J_{K+1}$ and the ones
given by the currents $J_i$ times the standard field combination
$C$,  contribute to the total fermionic current
already evaluated in the text and given in eq.(\ref{eq:13}). Therefore we can subtract them and
we are left with 
\be {\cal L}'=\frac 12\sum_{i=1}^{K+1}
f_i^2\left(\bar B_{i-1}+x_i C\right)^2-\sum_{i=1}^K J_i\bar V_i\,,\ee
where the fields $\bar V_i$ are the fields $V_i$ with the standard
model contribution subtracted, that is, using eq.(\ref{A15})
\be\bar V_i=\frac 1{g_i}\sum_{j=1}^K\theta_{j,i}\bar B_j\,.\ee We see
immediately that the terms linear in $C$ vanishes. In fact \be
\sum_{i=1}^{K+1}f_i^2x_i\bar
B_{i-1}C=f^2\left(\sum_{i=1}^{K+1}\bar B_{i-1}\right)C=0\ee due to
the identity satisfied by the fields $\bar B_i$ (see eq.(\ref{indep})). On the other
hand the term in $C^2$ gives rise to the $\tilde W$ and $\tilde Z$ masses since
\be \frac 1 2\sum_{i=1}^{K+1} f_i^2 x_i^2C^2=\frac 1 2 f^2
(\gt\Wt-\gpt \tilde Y)^2\ee from which
\be
{\tilde M}_{W}^2=f^2 \gt^2,~~~
{\tilde
M}_Z^2=\frac{\tilde M_W^2}{c_{\tilde\theta}^2}\,.
\ee
Therefore the
quartic term is obtained from
\be {\cal L}^{quart}_{eff}=\frac {1} {2}
\sum_{i=1}^{K+1} f_i^2\bar B_{i-1}^2-\sum_{i=1}^K J_i\bar V_i \,.
\ee
After substitution we get 
\be {\cal L}^{ quart}_{eff}=\frac
1{2f^2}\sum_{j,\ell=1}^K z_j z_\ell L_j L_\ell-\frac 12
\sum_{i,j,\ell=1}^K\frac 1{f_{i+1}^2}\theta_{i,j}\theta_{i,\ell}
L_jL_\ell\,,\ee or 
\be {\cal L}^{ quart}_{eff}=\frac
1{2f^2}\left(\sum_{i=1}^K z_i L_i\right)^2-\frac 1{2f^2}
\sum_{i=1}^K x_{i+1}\left(\sum_{j=1}^iL_j\right)^2\,.\ee 
Since \be
L_i=\frac {b_i}{1+\sum_{i=1}^K
b_i}\bar\psi_L\gamma^\mu\frac{\tau^a}2\psi_L\ee  we find
\be{\cal
L}^{quart}_{eff}
=\beta\sum_a\left(\bar\psi_L\gamma^\mu\frac{\tau^a}2\psi_L\right)^2
\ee
with
\be \beta=\frac 1{8f^2}\left(\bar b_K-b\right)^2-\frac 1
{8f^2}\sum_{i=1}^K x_{i+1}\bar b_i^2\ee
where \be
b=2\frac{\sum_{i=1}^Ky_i b_i}{1+\sum_{i=1}^K b_i},~~~\bar
b_i=2\frac{\sum_{j=1}^i b_j}{1+\sum_{j=1}^K b_j}\,.\ee

As an example let us consider the simple case \be
b_i=b_c,~~~~f_i=f_c\,.\ee It follows \be
f^2=\frac{f_c^2}{K+1},~~~x_i=\frac 1 {K+1},~~~y_i=\frac{i}{K+1}\ee
and \be \bar b_i=2\frac{i b_c}{1+Kb_c},~~~b=\frac{Kb_c}{1+Kb_c}\,.\ee
Therefore \be \beta=\frac 1{8f^2}\frac{K^2 b_c^2}{(1+K
b_c)^2}-\frac 1 {2f^2(K+1)}\frac{b_c^2}{(1+Kb_c)^2}\sum_{i=1}^K
i^2=-\frac 1{24 f^2}\frac{(Kb_c)^2}{(1+Kb_c)^2}\frac{K+2} K\ee or,
in terms of the parameter $b$, \be \beta=-\frac {b^2}{24
f^2}\frac{K+2}K\,.\ee


\end{document}